\documentclass[aps,pre,twocolumn,superscriptaddress,10pt]{revtex4-2}
\usepackage{amsmath}
\usepackage{hyperref}
\hypersetup{hidelinks}
\usepackage{physics}
\usepackage{bbm}
\usepackage{bbold}
\usepackage{graphicx}

\newcommand{\HH}{\mathcal{H}}
\newcommand{\FF}{\mathcal{F}}
\newcommand{\C}{\mathbb{C}}
\newcommand{\1}{\mathbb{1}}
\newcommand{\ee}{\mathrm{e}}
\newcommand{\ii}{\mathrm{i}}
\begin{document}

\author{{\L}ukasz Pawela}
\affiliation{Institute of Theoretical and Applied Informatics, Polish Academy of Sciences, Ba{\l}tycka 5, 44-100 Gliwice, Poland}
\email{lpawela@iitis.pl}
\title{Temporal-order-controlled Parrondo reversal in a periodically switched
quantum walk acting on NEQR image states}

\begin{abstract}
We study a coherent reversible map of novel enhanced quantum representation
(NEQR) image states, combining public permutations with a classically
controlled Parrondo discrete-time quantum walk on the color register.
Inversion retains every register coherently. An exact-expectation sweep with
an effect-size threshold identifies five cases in which two losing fixed-coin
walks form a winning switched walk. At the selected point, exhaustive
enumeration supplies a matched temporal-order control: with identical
angles, length, initialization, and coin counts, \(00111\) yields
\(\Delta_C=+7.21\), whereas \(01101\) yields \(\Delta_C=-8.96\).
Translation covariance makes each conditional output-color distribution a
shifted copy of one exact walk kernel. Exact diagnostics expose finite-shot
artifacts and repeated-copy relative-shift leakage. They demonstrate an
order effect, but neither a reversal-caused scrambling advantage nor
cryptographic security. We additionally quantify finite-shot convergence,
wrong-control sensitivity, and a depolarizing-noise
basis-recovery bound. Resource estimates show that arbitrary NEQR preparation
dominates the \(O(q^2+qr)\) walk layer, delimiting the scope of the
construction.
\end{abstract}


\maketitle

\section{Introduction}

Quantum image processing uses features such as superposition and entanglement
in high-dimensional Hilbert spaces to represent and manipulate visual
information~\cite{abd2019encryption}. Once an image is stored as a quantum
state, a natural primitive is a controlled reversible transformation that
redistributes computational-basis measurement outcomes while remaining
exactly invertible before measurement. Constructing such a map consistently,
and determining what information remains accessible under repeated
measurements, is the subject of this work. The transmission of classical
ciphertext images and indistinguishability-based encryption lie outside its
scope.

Quantum image encryption schemes now commonly build on efficient quantum
image representations such as the Novel Enhanced Quantum
Representation (NEQR)~\cite{zhang2013neqr}, which gives deterministic
retrieval and pixel-level addressability of grayscale
images~\cite{abd2019encryption}. One line of work uses quantum walks, the
quantum analogues of classical random walks, whose dynamics combine
superposition, ballistic spreading, and entanglement generation. Discrete-time
quantum walks
(DTQWs)~\cite{Aharonov2001,AaronsonAmbainisToC2005,lovett2010universal}
(see~\cite{VenegasAndraca2012} for a review) suit quantum circuit
implementation because of their discrete structure, and appear in quantum
algorithms and cryptographic
protocols~\cite{razzoli2024efficient,VenegasAndraca2012,Portugal2018,Kadian2021}.

In quantum image encryption, DTQWs have often been proposed as a source of
complex masks or keys. For example, Ref.~\cite{abd2019encryption} uses a
quantum walk on a cycle in connection with NEQR images. Such constructions must
be handled with care: a pixelwise classical XOR operation is not automatically
obtained by applying gates transversally to two independently addressed quantum
images, and discarding a quantum key image after it has become entangled with
the data generally destroys information needed for
decryption~\cite{grassl2020comment}. Recent quantum-walk image-encryption work
also combines walk-generated keys with explicit scrambling and XOR diffusion
circuits~\cite{zeng2026entropy}. Its transmitted cipher-image metrics belong
to a different statistical object from the coherent state studied here, which
is why semiclassical mask-generation schemes, measured diagnostic images, and
coherent scrambled quantum states need to be kept apart.

The present work introduces no separate quantum key image. We
study a coherent, controlled unitary transformation of an NEQR image state. The
classical control vector specifies the Parrondo coin sequence, coin parameters,
walk length, and color offset. The spatial and position--color layers used in
the present experiments are fixed and public. The transformed object is the full
quantum state produced by this unitary transformation. Applying the inverse
unitary with the same control parameters restores the original image state before
measurement. Conversely, measuring the scrambled registers before inversion
produces only a classical sample from the scrambled state's computational-basis
distribution and is not, by itself, a decryptable coherent state. In the
experiments below, such measurements are used as diagnostics of scrambling.

Three design classes are sometimes conflated in quantum image encryption. The
first class attempts to build a quantum key image and combine it with the data image
by a pixelwise XOR-like operation. Such constructions require special care
because discarding or measuring a key register after it has become entangled
with the data can destroy information needed for deterministic decryption. The
second class uses a quantum walk only to generate a sampled classical mask or
keystream. This can define a conventional classical image-encryption workflow,
but the walk is then used as a quantum or quantum-inspired source of
randomness. The third class, adopted here, keeps the entire transformation
coherent and decrypts by applying the inverse keyed unitary before readout.
Table~\ref{tab:scheme-comparison} summarizes this distinction.

\begin{table*}[!htp]
\begin{tabular}{p{0.21\textwidth}p{0.25\textwidth}p{0.25\textwidth}p{0.20\textwidth}}
\hline
Design & Classical object used for diagnostics & Inversion object & Main limitation \\
\hline
Sampled DTQW mask schemes & A measured walk distribution is converted into a
classical image mask or scrambling pattern. & The receiver must reproduce the
same classical mask and invert the corresponding classical operations. &
Measurement of the walk distribution is part of the cryptographic object, so
the analysis is tied to the sampled mask. \\
Independent quantum key-image constructions & A second quantum image or key
register is combined with the data image. & Correct inversion requires that
all registers carrying information about the operation remain coherent and
available. & If an entangled key image is discarded or measured prematurely,
the remaining state need not contain enough information for deterministic
decryption. \\
Present coherent keyed-unitary construction & Classical diagnostics are
obtained only by measuring the coherent scrambled state after the keyed unitary
has been applied. & Inversion is \(U_K^\dagger\) applied to the full coherent
state, including the coin register, before readout. & The reported image
statistics are scrambling diagnostics, not a stand-alone security proof for a
classical ciphertext-image protocol. \\
\hline
\end{tabular}
\caption{Conceptual comparison between the present construction and two common
quantum-walk image-encryption design patterns. The distinguishing feature of
the present approach is that the quantum walk is used as one reversible layer
inside a single keyed unitary, rather than as a measured mask or as a discarded
quantum key image.}
\label{tab:scheme-comparison}
\end{table*}

Our scheme integrates three reversible quantum layers. Spatial diffusion is a
permutation layer that scatters pixel information globally across the image
lattice. Position-color mixing is a nonlinear Boolean position--color
operation, built from Toffoli gates, that changes local correlations and plays
the role of classical confusion. DTQW substitution is a unitary Parrondo
quantum walk on the color register, driven by a time-dependent coin sequence
selected by the classical key.
Reversible spatial scrambling of quantum images has also been studied in its
own right, through quantum realizations of the Arnold, Fibonacci, and
Hilbert transforms~\cite{jiang2014arnold,jiang2014hilbert}. The bit-reversal
permutation used here is a deliberately simple \(O(n)\) member of this
family, and any of these keyed scrambling permutations could be substituted
for it.

The DTQW formalism involves a Hilbert space that is the tensor product of a
position register and an internal coin register. Each time step consists of a
quantum coin operation followed by a conditional shift, efficiently implemented
in quantum circuits via the quantum Fourier transform (QFT) for diagonalizing
the conditional shift operator~\cite{razzoli2024efficient}. The periodic
dynamics, recurrence, and repeated generation of entangled states exhibited by
quantum walks bear directly on coherent scrambling
applications~\cite{razzoli2024efficient,Panda2023}.

The Parrondo
paradox~\cite{Parrondo1996,Parrondo2000,Abbott2010,MeyerBlumer2002}, originally
developed in game theory and statistical physics, describes the counterintuitive
phenomenon where alternating between two losing strategies results in a winning
outcome. Quantum analogues of Parrondo's games, especially those involving
quantum walks, display richer behavior owing to quantum coherence and
entanglement~\cite{chandrashekar2011parrondo,flitney2012,pawela2013cooperative,Walczak2021,Lai2020,rajendran2018implementing,rajendran2018playing,pires2020timedependent,walczak2022threecoins,walczak2024shift,panda2022entangled,mittal2024inhomogeneous,jan2020experimental,trautmann2022momentum,jan2023territories}.
In particular, the paradox
manifests in quantum walks on cycles and lines, where switching between quantum
coin operations that individually produce a ``losing'' outcome can, when
combined, yield a ``winning''
result~\cite{pawela2013cooperative,Walczak2021,rajendran2018implementing}.
The effect has been demonstrated experimentally with two rotation
operators~\cite{jan2020experimental}, analyzed in momentum-space
implementations~\cite{trautmann2022momentum}, and mapped across sequence
periods and initial relative phases for two-state
coins~\cite{jan2023territories}. The narrower question here is whether the
spatially homogeneous, phaseless rotations used by the QFT cycle circuit show
a threshold-qualified reversal and whether that reversal persists over a
long pre-wrap numerical window.
The closest use of Parrondo dynamics in a cryptographic context appears to be
the chaotic switching of quantum-coin Parrondo games as a bitstream generator
proposed in Ref.~\cite{lai2021chaotic}. In contrast, the present work embeds
the Parrondo walk as one reversible layer of a coherent keyed unitary acting
directly on a quantum image state.

For the cycle walk used here, the paradox criterion of two individually
losing games combining into a winning one is made operational through the
mean displacement \(\Delta=\mathbb{E}[X_r]-\delta\) after \(r\) steps, where
\(\delta\) is the initial color-register position. The two fixed games use
\(\hat C_0\) and \(\hat C_1\) at every step, while the combined game uses the
key sequence \(m\) to switch between these coins. Because the coin rotations
and conditional shifts do not commute, the combined product is not a convex
average of the two fixed-coin evolutions: the sequence changes the relative
phases accumulated by the two propagation directions and can therefore change
the sign of the final drift. We call a parameter set paradoxical when
\(\Delta_A<-\tau\), \(\Delta_B<-\tau\), and \(\Delta_C>\tau\) for an explicit
effect-size threshold \(\tau\), with all displacements computed from exact
circuit statevectors rather than from finite samples.

A sampled displacement of a ballistically spreading walk carries a standard error of order
\(\sigma/\sqrt{S}\), which reaches \(\approx 3.1\) positions at a reference
budget of \(S=1024\) shots for degenerate parameter sets, where
\(\sigma=r\). A plain sign test on sampled displacements therefore
flags a ``paradox'' with probability up to \(1/8\) even for parameter sets
whose three games are physically identical, such as coin angles for which the
coin operator degenerates to the identity. Requiring a margin
\(\tau\) applied to exact expectations removes this degeneracy. The margin is
a deterministic selection rule rather than a statistical significance test. We also
find that for the phaseless rotation coins used here the simple alternation
\(m=0101\ldots\) never produces a reversal passing the threshold anywhere on
the parameter grid. Period-asymmetric repeated-coin sequences, such as
\(m=001\,001\ldots\), \(0011\,0011\ldots\), and
\(00111\,00111\ldots\) (respectively \(AAB\), \(AABB\), and \(AABBB\) when
\(m_\ell=0\) selects \(\hat C_0\)), are also tested. On our grid the selected
paradox rows arise only from the \(001\) and \(00111\) periods. In this
manuscript the strongest reversal in that preset sweep fixes the angles and coin composition for a
second, exhaustive ordering experiment. Enumerating all ten length-five words
with two uses of \(\hat C_0\) and three of \(\hat C_1\) identifies a
threshold-losing word outside the winning word's cyclic-rotation class. This
strict control changes only temporal order, and the starting phase enters
explicitly as part of the ordered control.

The contributions are as follows. A consistent, fully coherent reversible
formulation of a quantum-walk image transformation avoids the
discarded-key-image and measured-mask pitfalls. A methodological correction
for Parrondo classification in quantum walks replaces sampled sign tests by
exact-expectation displacements with an explicit effect-size threshold, shown
below to be necessary to avoid spurious paradox classifications at realistic
shot budgets, and adds a same-composition intervention that isolates temporal
order at the selected point. An exact translation-covariance analysis of the
image layer then makes explicit what finite-shot image diagnostics can and
cannot certify. What emerges is a threshold-qualified, long-window Parrondo
reversal in the QFT-implemented cycle walk whose interpretation as an
image-transformation advantage is sharply limited.

\section{Quantum walk implementation}

Quantum walks (QWs) generalize classical random walks and underlie a range of
quantum algorithms, simulation schemes, and cryptographic protocols. Among QW
models, the DTQW suits circuit implementation on quantum computers because it
discretizes both time and space~\cite{razzoli2024efficient}.

\subsection{Discrete-time quantum walk on a cycle}

In the DTQW, the walker has two degrees of freedom: position and an internal
two-level coin. Its dynamics are described by a complex Euclidean space that is
the tensor product of the coin and position spaces, so the whole evolution
takes place in the space
\begin{equation}
\HH = \HH_C \otimes \HH_P = \C^2 \otimes \C^N,
\end{equation}
where $N$ is the number of positions on the cycle~\cite{Aharonov2001,VenegasAndraca2012}.

Each step of the walk is described by the unitary operator
\begin{equation}
    U = S ( C \otimes \1_{\HH_P} ),
\end{equation}
where $C \in U(\C^2)$ is a unitary coin operator, $\1_{\HH_P}$ is the
identity on the position space, and $S$ is the conditional shift operator. The operator
$S$ moves the walker clockwise or counterclockwise around the cycle, conditioned
on the coin state:
\begin{equation}
    S\ket{s_C}\ket{j_P} = \ket{s_C}\ket{[(j + 2s - 1) \bmod N]_P},
\end{equation}
where $s \in \{0,1\}$ and $j \in \{0,\ldots,N-1\}$.

\subsection{Quantum circuit design} \label{sec:walk-circuit}

Circuit efficiency limits what DTQWs can do on current quantum hardware. The
design of~\cite{razzoli2024efficient} reduces multi-qubit gate count and
circuit depth relative to earlier approaches. For a DTQW on an $N=2^n$-cycle over $t$ steps,
their scheme requires only $O(n^2 + n t)$ two-qubit gates, as opposed to the
$O(n^2 t)$ scaling of previous QFT-based schemes.

That design diagonalizes the conditional shift operator using the quantum
Fourier transform (QFT) without swap gates, $\tilde{\FF}$. Applying
$\tilde{\FF}$ to the position register reduces each time step, in the Fourier
basis, to a layer of single-qubit and controlled phase gates acting
conditionally on the coin state, applied together with the coin operator. The
inverse QFT then returns the position register to the computational basis.
The resulting operator for $t$ steps is
\begin{equation}
    U^t = (\1_C \otimes \tilde{\FF}^\dagger) \left[ \Sigma (C \otimes \1_P) \right]^t (\1_C \otimes \tilde{\FF}),
\end{equation}
where $\Sigma$ is a diagonal operator in the Fourier basis, applying different
phase shifts to the position states depending on the coin
state. It has the form:
\begin{equation}
    \Sigma = \ketbra{0_C}{0_C} \otimes \Omega^\dagger + \ketbra{1_C}{1_C} \otimes \Omega,
\end{equation}
and
\begin{equation}
    \Omega = \mathrm{diag}(1, \ee^{2\pi \ii / N}, \ee^{4\pi \ii / N}, \ldots, \ee^{2\pi \ii (N-1) / N}) = \bigotimes_{k=1}^{n} R_k,
\end{equation}
with
\begin{equation}
    R_k = \begin{pmatrix}
        1 & 0 \\
        0 & \ee^{2\pi \ii / 2^k}
    \end{pmatrix}.\label{eq:rk}
\end{equation}

\begin{figure}[!htp]
\centering
\includegraphics[width=\columnwidth]{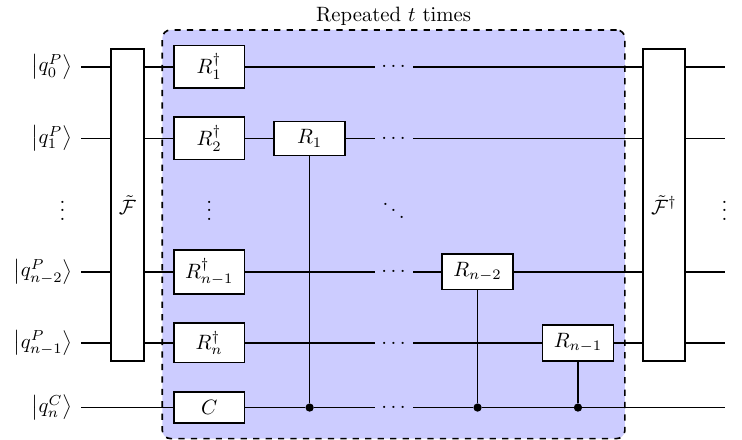}
\caption{Quantum circuit for a discrete-time quantum walk on a cycle with $n$
position qubits. The quantum Fourier transform $\tilde{\FF}$ and its inverse do
not contain the SWAP operations (see~\cite{QiskitQFT} for details). The
operations $R_k$ are defined in Eq.~\eqref{eq:rk}.}\label{fig:walk-circuit}
\end{figure}

Fig.~\ref{fig:walk-circuit} shows the circuit, illustrating that only one QFT
and one inverse QFT are needed, regardless of $t$, with the central block
repeated for each time step. This approach can be further simplified for an
initially localized walker (in the state $\ket{\phi_C}\ket{0_P}$) by replacing
the QFT with a layer of Hadamard gates (see Appendix B
in~\cite{razzoli2024efficient}).

\section{Coherent quantum image scrambling based on quantum walks}

This section specifies the coherent scrambling map studied here. A classical
control vector selects the walk inside a reversible unitary acting on the NEQR
image and an explicit coin qubit, with no separate quantum key image and no
measurement of a quantum walk to obtain a classical mask. We use key notation
for this control vector without assigning it a cryptographic security level.

\subsection{Image state and keyed walk}

Let \(q=8\) be the number of color qubits. A grayscale image of size
\(2^n \times 2^n\) is encoded in the NEQR form
\begin{equation}
    \ket{I} =
    \frac{1}{2^n}
    \sum_{y=0}^{2^n-1}\sum_{x=0}^{2^n-1}
    \ket{c_{y,x}}_{\mathrm{col}}\ket{y}_Y\ket{x}_X
    \in
    \C^{2^q} \otimes \C^{2^n} \otimes \C^{2^n},
    \label{eq:neqr-state}
\end{equation}
where \(\ket{c_{y,x}}_{\mathrm{col}}\) is the computational-basis encoding of
the 8-bit grayscale value at pixel \((x,y)\). The additional coin register is a
single qubit, so the full working space is
\begin{equation}
    \HH_{\mathrm{coin}}\otimes\HH_{\mathrm{col}}\otimes\HH_Y\otimes\HH_X
    =
    \C^2\otimes\C^{2^q}\otimes\C^{2^n}\otimes\C^{2^n}.
\end{equation}

The classical walk-control vector is denoted
\begin{equation}
    K = (m,r,\theta_1,\theta_2,\delta),
\end{equation}
where \(m\) is a binary coin sequence of length \(r\), \(r\) is the number of
walk steps, \(\theta_1,\theta_2\) parameterize the two coins, \(\delta\) is an
optional bitwise offset applied to the color register before the walk. We use
\(\pi\) and \(\mu\) separately for the reversible spatial permutation and
position--color mixing maps. In the parameter sweep and image-scrambling
experiments, \(\delta=127\) and \(\pi,\mu\) are fixed public circuit
parameters, and only the Parrondo parameters are varied. The later sensitivity
study perturbs \(\delta\). The notation \(U_K\) suppresses the
fixed public dependence on \(\pi\) and \(\mu\).

The two coin operators are the real rotations
\begin{align}
\hat{C}_0 &= \begin{pmatrix}
\cos \theta_1 & -\sin \theta_1 \\
\sin \theta_1 & \cos \theta_1
\end{pmatrix}, \\
\hat{C}_1 &= \begin{pmatrix}
\cos \theta_2 & -\sin \theta_2 \\
\sin \theta_2 & \cos \theta_2
\end{pmatrix},
\end{align}
implemented as \(U(2\theta_i,0,0)\) gates. The coin
\(\hat{C}_i\) becomes trivial (proportional to the identity) at
\(\theta_i\in\{0,\pi\}\). Such degenerate points are excluded by the
effect-size criterion of the parameter sweep. At step \(\ell\), the applied
coin is \(\hat{C}_0\) if \(m_\ell=0\) and \(\hat{C}_1\) if \(m_\ell=1\). Since
\(m\) has length \(r\), a coin is defined for every step. Together with the
conditional shift, these choices define a unitary walk \(W_K\) on
\(\HH_{\mathrm{coin}}\otimes\HH_{\mathrm{col}}\), where the color register is
the walker's position register on a \(2^q\)-cycle. In the implementation,
\(\delta\) is realized by applying \(X\) gates to the color bits set in
\(\delta\) before the QFT-based walk block. On a general color state this
implements the bitwise offset \(\ket{c}\mapsto\ket{c\oplus\delta}\). The walk
block also begins with a Hadamard gate on the coin register, so the walk
effectively starts from the real equal-weight coin state
\(\ket{+}=(\ket{0}+\ket{1})/\sqrt{2}\). The paradox classification below
refers to this initialization, to which the effect is sensitive: for the
complex unbiased state \((\ket{0}+\ii\ket{1})/\sqrt{2}\), a symmetry of
real-coin walks forces every displacement to vanish identically, so the
drifts of the fixed games and of the combined game alike originate in the
interference between the real initial coin state and the real coin
rotations. The
circuit realization of this walk is the QFT-based construction of
Section~\ref{sec:walk-circuit}.

\subsection{Scrambling procedure}

The coherent scrambling map is a keyed unitary \(U_K\). Starting from
\begin{equation}
    \ket{\Psi_0} = \ket{0}_{\mathrm{coin}}\otimes\ket{I},
\end{equation}
we apply three reversible layers.

Spatial diffusion applies a reversible permutation
\(U_{\mathrm{diff}}^{(\pi)}\) to the position registers, implemented in the
simulations by bit reversal on \(X\) and \(Y\) followed by a swap of the two
coordinate registers,
\begin{equation}
    \ket{\Psi_1} =
    (\1_{\mathrm{coin,col}}\otimes U_{\mathrm{diff}}^{(\pi)})
    \ket{\Psi_0}.
\end{equation}
Position-color mixing then applies a reversible mixing unitary
\(U_{\mathrm{conf}}^{(\mu)}\) whose CNOT and Toffoli gates couple coordinate
bits to color bits,
\begin{equation}
    \ket{\Psi_2} = U_{\mathrm{conf}}^{(\mu)}\ket{\Psi_1}.
\end{equation}
DTQW substitution finally applies the keyed Parrondo walk to the coin and
color registers,
\begin{equation}
    \ket{\Psi_{\mathrm{enc}}} =
    (W_K\otimes\1_{YX})\ket{\Psi_2}.
\end{equation}

Equivalently,
\begin{align}
    \ket{\Psi_{\mathrm{enc}}}
    &= U_K\ket{\Psi_0}, \\
    U_K
    &= (W_K\otimes\1_{YX})
    U_{\mathrm{conf}}^{(\mu)}
    (\1_{\mathrm{coin,col}}\otimes U_{\mathrm{diff}}^{(\pi)}).
    \label{eq:keyed-unitary}
\end{align}
The coherent state \(\ket{\Psi_{\mathrm{enc}}}\), including the coin register,
is the scrambled quantum image state in this model.

\subsection{Inversion and diagnostic readout}

Inversion requires the coherent state \(\ket{\Psi_{\mathrm{enc}}}\) and the
same classical key \(K\). The receiver applies
\begin{equation}
    U_K^\dagger =
    (\1_{\mathrm{coin,col}}\otimes U_{\mathrm{diff}}^{(\pi)\dagger})
    U_{\mathrm{conf}}^{(\mu)\dagger}
    (W_K^\dagger\otimes\1_{YX})
\end{equation}
to recover
\begin{equation}
    U_K^\dagger\ket{\Psi_{\mathrm{enc}}} = \ket{\Psi_0}.
\end{equation}
The NEQR image may then be read out. If a complete classical image is desired,
the readout requires repeated preparations or an equivalent NEQR retrieval
procedure to obtain all pixel values.

By contrast, measuring the scrambled registers before applying \(U_K^\dagger\)
produces only classical samples from the distribution induced by
\(\ket{\Psi_{\mathrm{enc}}}\). Those samples are not a coherent quantum
state and are not expected to be invertible by the inverse quantum
circuit. In the numerical section, the displayed diagnostic reconstructions
come from repeated measurements of independently prepared scrambled states. Only the color and coordinate registers are measured. The
coin register is left unmeasured and is thereby marginalized over, so the
diagnostics refer to the reduced computational-basis distribution of the image
registers and involve no post-selection on the coin. The samples are used to
assess scrambling through correlations, histograms, entropy, NPCR, and UACI,
not to define a classical ciphertext-image transmission protocol.

\subsection{High-level resource estimates}

The coherent construction acts on
\begin{equation}
    Q = 1 + q + 2n
\end{equation}
qubits for a \(2^n\times2^n\) grayscale image with \(q\) color qubits: one coin
qubit, \(q\) color qubits, and two \(n\)-qubit coordinate registers. For the
\(64\times64\) images used below, \(n=6\) and \(q=8\), so the ideal circuit uses
21 quantum qubits and measures 20 image-register bits. The estimates in
Table~\ref{tab:resources} count the high-level operations appearing in the
implemented circuit before decomposition into a specific hardware basis.

\begin{table*}[!htp]
\begin{tabular}{p{0.19\textwidth}p{0.37\textwidth}p{0.18\textwidth}p{0.18\textwidth}}
\hline
Circuit component & High-level resources & Scaling & Count for \(64\times64\), \(q=8\), \(r=100\) \\
\hline
NEQR loading & \(2n\) Hadamard gates for the coordinate superposition, followed
by one \(2n\)-controlled \(X\) gate for each nonzero color bit in the classical
image. & \(O(q4^n)\) multi-controlled gates in the worst case. & \(12\)
Hadamards and at most \(32768\) \(C^{12}X\) gates. The actual count depends on
the image bit pattern. \\
Spatial diffusion & Bit reversal on both coordinate registers and a swap of
the two coordinates. & \(O(n)\) swaps. & \(2\lfloor n/2\rfloor+n=12\) swaps. \\
Position-color mixing & For \(L=\min(n,q)\), the implemented layer uses
\(4L\) CNOT gates and \(L\) Toffoli gates. & \(O(\min(n,q))\). & \(24\) CNOT
gates and \(6\) Toffoli gates. \\
DTQW substitution & \(\mathrm{wt}(\delta)\) initial \(X\) gates implementing
the bitwise offset \(c\mapsto c\oplus\delta\), one coin
Hadamard, one QFT and inverse QFT on the \(q\)-qubit color register, and \(r\)
walk steps. Each step has one coin rotation, \(q\) phase gates, and \(q-1\)
controlled-phase gates. & \(O(q^2+qr)\) one- and two-qubit walk gates. & \(7\)
\(X\) gates, \(17\) Hadamard gates including the QFT pair, \(100\) coin
rotations, \(800\) phase gates, and \(756\) controlled-phase gates. \\
Diagnostic readout & Measurement of the color and coordinate registers. &
\(O(q+n)\). & \(20\) measured bits. \\
\hline
\end{tabular}
\caption{High-level resource estimates for the implemented coherent scrambling
circuit. Counts are expressed in the gate set used to build the circuit, before
decomposing multi-controlled \(X\), Toffoli, SWAP, and controlled-phase gates
into a device-native basis. The NEQR loading term dominates for arbitrary
classical input images, while the QFT-based DTQW layer has the milder
\(O(q^2+qr)\) scaling discussed in Section~\ref{sec:walk-circuit}.}
\label{tab:resources}
\end{table*}

Two costs of very different size enter this estimate. The
Parrondo DTQW layer is compact once the image is already present as an NEQR
state, especially because the QFT and inverse QFT are applied only once.
Loading an arbitrary classical image into NEQR form, by contrast, is expensive
and dominates the high-level gate count. The present construction is
accordingly best read as a coherent scrambling layer for quantum image states,
or for settings where image preparation is amortized or supplied by a preceding
quantum process, rather than as an immediately practical replacement for
classical image encryption.

\subsection{Scope and limitations}

The construction studied here is a coherent quantum image scrambling map
rather than a complete classical image cipher. Inversion requires the full
coherent state \(\ket{\Psi_{\mathrm{enc}}}\), including the coin register,
together with the classical key \(K\), and measurements before inversion
yield only classical samples.

The image quantities reported below are accordingly descriptive rather than
cryptographic. Low correlations, high modal entropy, and large NPCR/UACI can
characterize a chosen reconstruction estimator, but they neither establish
uniform conditional distributions nor constitute a security proof against
chosen-plaintext, known-plaintext, or quantum adversarial attacks.
A unitary map cannot increase the distinguishability of its inputs. Two NEQR states of images differing in a
single pixel have overlap at least \(1-4^{-n}\), and the scrambled states
retain exactly the same overlap. The classical differential ``avalanche''
measured by NPCR/UACI in sampled-mask schemes is a property of the
measurement-and-mask step, not something a coherent unitary layer provides at
the state level. Plaintext sensitivity in this
construction resides in the statistics of measurement outcomes, not in the
geometry of the states.

Two further consequences follow, both under access to independently prepared
copies through repeated uses of the map rather than possession of a single
unknown state, which cannot be cloned. First, the map is deterministic. The
control vector contains no per-message randomizer, so fixed parameters
and plaintext always produce the same scrambled state. Combined with exact
overlap preservation, this means that an adversary holding several scrambled
states can estimate their pairwise overlaps with SWAP tests, learning how
similar the underlying plaintexts are and detecting repeated plaintexts
outright. The present deterministic map cannot hide equality or input-state
overlap under repeated reuse of the same fixed control vector.
Avoiding this leakage would require an appropriate per-message randomization
and security model, which we do not attempt. Second, the
implemented parameter family is not proposed as a security parameter. The
diffusion and mixing layers are fixed and public, so all variable content
resides in the walk block. The experiment uses four preset periodic sequences
rather than \(r\) independently sampled secret bits, and no discretization or
prior distribution is specified for the continuous coin angles. The width of
a recovery-fidelity curve is not a measure of key entropy or
parameter-estimation hardness. The offset contributes \(q\) bits and
orthogonalizes the coherent state, but provides no concealment of the measured
reconstruction (Section~\ref{sec:key-sensitivity}). No key size is quoted
here. A cryptographic instantiation would need randomized, keyed diffusion and
mixing families and a quantitative key-recovery analysis under explicit access
assumptions, neither of which is attempted.

Recovering a complete classical image from an NEQR state also requires
repeated preparations or an equivalent readout procedure, and exact preparation
of arbitrary NEQR images may dominate the resource cost, as made explicit in
Table~\ref{tab:resources}. Section~\ref{sec:key-sensitivity} quantifies the
sensitivity of the coherent map to key perturbations, and
Section~\ref{sec:noise} quantifies the effect of depolarizing gate noise in
the keyed walk layer on the scrambling/inversion cycle. Noise in the NEQR
preparation itself, fault-tolerant overheads, and hardware-specific
compilation costs are left for subsequent analysis.

A classical pixel perturbation made before NEQR preparation simply defines a
different input state. In the ideal circuit, \(U_K^\dagger U_K=I\) returns
that perturbed state exactly: the map neither amplifies nor removes the input
perturbation and is not a denoising procedure. This input-data issue is
distinct from imperfect state preparation or noisy implementation,
which are physical channels and require separate models.

\subsection{Experimental environment and simulation process}

All numerical results were generated with the Python/Qiskit implementation
included with the manuscript sources. The repository-level \texttt{env.yml}
file specifies Python 3.12 together with Qiskit 1.4.2, Qiskit Aer 0.16.0,
Qiskit IBM Runtime 0.38.0, NumPy 2.2.5, Pillow 11.2.1,
and Matplotlib 3.10.1. The results reported here use exact statevector
computations for the parameter sweep and the key-sensitivity study, direct
position-basis evolution for the exact kernels, and seeded multinomial
sampling from the exact conditional distributions for the finite-shot
visualizations. The latter is statistically equivalent to measuring the full
image circuit because of Eq.~\eqref{eq:conditional-shift}, and the independent
Qiskit Aer statevector route is retained as a circuit-level validation. We use
the Aer density-matrix simulator for the noise study and the Qiskit transpiler at
optimization level 1, except in the noise study, which transpiles at
optimization level 0 so that no gates are cancelled across the
scrambling/inversion boundary and the noisy gate count is preserved. No remote quantum-hardware runs are used for the tables
below. All sampled Aer simulator and transpiler calls use the base seed
\texttt{PARRONDO\_SEED=1234}, and the modal-convergence study uses 16
deterministically derived replicate seeds per budget. The main executable scripts are
\texttt{src/check\_paradox.py}, \texttt{src/matched\_control.py},
\texttt{src/conditional\_image\_diagnostics.py},
\texttt{src/compare\_quality.py},
\texttt{src/encrypt\_v2.py}, \texttt{src/per\_layer\_study.py},
\texttt{src/state\_level\_diagnostics.py},
\texttt{src/key\_sensitivity.py}, \texttt{src/noise\_robustness.py},
\texttt{src/asymptotic\_check.py}, and
\texttt{src/walk\_figure.py}. The whole pipeline is reproduced by
\texttt{src/launch.sh}.

The simulation has two stages. First, \texttt{src/check\_paradox.py} performs
the reduced-walk parameter sweep using an 8-qubit color walk, \(r=100\), and
initial position \(\delta=127\). The coin sequences are the periodic presets
\(m=0101\ldots\), \(001\,001\ldots\), \(0011\,0011\ldots\), and
\(00111\,00111\ldots\), and each coin angle \(\theta_1,\theta_2\) ranges over
the 41-point grid \(\{k\pi/40 : k=0,\ldots,40\}\). For every parameter set the
displacements \(\Delta_A\), \(\Delta_B\), and \(\Delta_C\) of the two fixed-coin
walks and the combined walk are computed from exact circuit statevectors, so the
classification contains no sampling noise. A parameter set is classified as
paradoxical when \(\Delta_A<-\tau\), \(\Delta_B<-\tau\), and \(\Delta_C>\tau\)
with \(\tau=5\). This deterministic effect-size filter excludes parameter
sets whose displacements vanish identically (for instance trivial coins at
\(\theta_i\in\{0,\pi\}\)). For scale, \(\tau\) is 1.6 times the largest
\(1024\)-shot standard error of the displacement estimator across the sweep
(\(\approx 3.1\) positions, attained at degenerate coins where \(\sigma=r\)).
The filter is not a hypothesis test: it attaches no \(p\)-value to a grid
point and applies no correction for screening multiple parameter sets. Of the \(6724\)
parameter sets in
the sweep, five satisfy the criterion, one with the \(001\) period and four
with the \(00111\) period, while the balanced \(0011\) period and the
alternating sequence \(0101\ldots\) yield none.

\begin{figure*}[!htp]
\centering
\includegraphics[width=\textwidth]{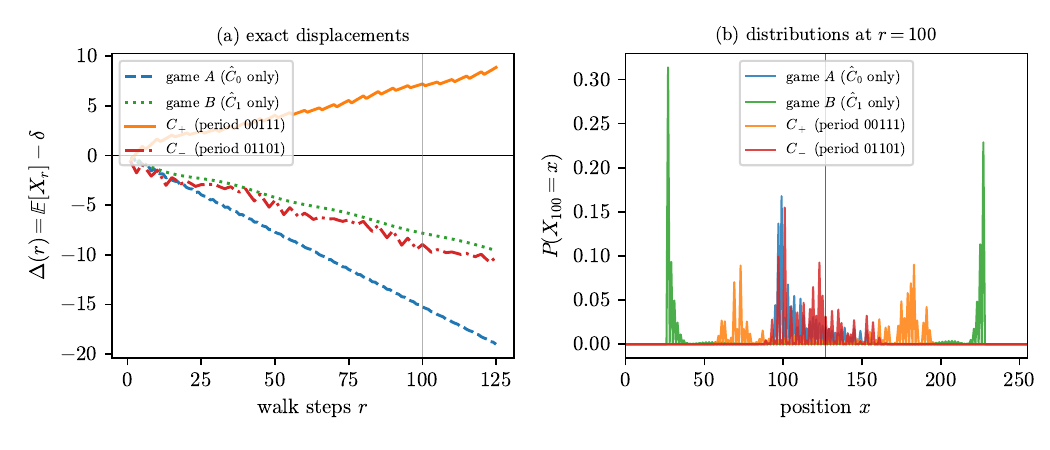}
\caption{Exact walk diagnostics for the matched comparison
(\(\theta_1=3\pi/5\), \(\theta_2=39\pi/40\), \(\delta=127\)).
(a) Exact mean displacement \(\Delta(r)\) of the two fixed-coin games, the
reversal word \(C_+=00111\), and the matched losing word \(C_-=01101\), as a
function of the number of walk steps. The vertical line marks \(r=100\) used
in the image simulations. The
curves are shown for \(r\le125<N/2\), the domain in which the ballistic
components have not wrapped around the cycle and the linear displacement
measures drift. (b)
Exact position distributions of all four games after \(r=100\) steps, with
the vertical line marking the initial position \(\delta=127\). Both fixed games and
\(C_-\) drift backward, whereas \(C_+\) drifts forward.}
\label{fig:walk-displacement}
\end{figure*}

These parameter choices are tied to the implementation rather than optimized as
security parameters. The walk size \(N=2^8\) matches the 8-bit grayscale color
register: the DTQW acts on color values, not on the \(64\times64\) spatial
lattice. The value \(r=100\) lies well inside the window in which the selected
paradoxical regime keeps its sign structure: the signs
\(\Delta_A,\Delta_B<0<\Delta_C\) hold at every integer step count
\(3\le r\le125\), throughout
the domain in which the displacement on the cycle measures drift
(walks from \(\delta=127\) begin to wrap around the cycle at
\(r\approx N/2=128\), where the linear mean loses its meaning as a game
outcome), and all three displacements exceed the margin \(\tau\) at every
integer \(72\le r\le 125\). The margin also holds at \(r=69\) and \(70\),
with a single dip of the combined game to \(\Delta_C=4.91\), just below
\(\tau\), at \(r=71\) (see Fig.~\ref{fig:walk-displacement}). The choice also
keeps the full image simulation tractable under the diagnostic shot budget
used below. The 41-point
angle grid is a reproducible uniform screen over \([0,\pi]\) whose resolution
\(\pi/40\) resolves five qualifying rows in the tested presets. This is not a
grid-convergence study, and the screen may miss narrower regions or off-grid
points. The selected reversal is best read as a representative
point from this fixed screen, not as an optimized cryptographic parameter
search.

We construct its control with \texttt{src/matched\_control.py}. Starting from
the threshold-qualified row with the largest \(\Delta_C\), whose shortest
period is \(00111\), we retain the angles, \(N\), \(r\), \(\delta\), and
initial coin state \(\ket{+}\), and enumerate all
\(\binom{5}{2}=10\) period words containing two uses of \(\hat C_0\) and
three of \(\hat C_1\). Because \(r=100\) is a multiple of five, every word
uses the coins exactly 40 and 60 times. Starting phase can change a
finite-time ordered product, so it is treated as part of the intervention:
all cyclic rotations are reported, and the primary losing control is required
to lie outside the reversal word's cyclic-rotation class. The only word
satisfying that requirement, \(01101\), is threshold-losing and differs from
\(00111\) by swapping positions two and four within each period, so that over
20 periods 40 of the 100 time-step labels differ. It also preserves
the first and last symbols. The
primary pair therefore isolates temporal order at this fixed parameter point.

Second, \texttt{src/conditional\_image\_diagnostics.py} applies the two
selected orderings to four \(64\times64\) grayscale images. The public
spatial-diffusion and position--color maps determine the walk's input color at
each output coordinate. Equation~\eqref{eq:conditional-shift} then gives its
exact conditional output distribution as a shifted copy of the selected walk
kernel. We draw 250000 seeded multinomial samples from this joint
distribution and assign the most frequently observed color at each coordinate
to form the diagnostic reconstruction, breaking finite-sample ties uniformly.
The third row is an analytic ideal-inversion reference: because the modeled
map is exactly unitary, \(U_K^\dagger U_K=\mathbbm{1}\), the generator renders
that row directly from the input rather than treating it as a second
experiment. The slower full-circuit statevector route in
\texttt{src/compare\_quality.py} explicitly composes the implemented inverse,
recovers the input, gives the same conditional probabilities, and is
retained as an independent implementation check.

\begin{table*}[!htp]
\centering
\begin{tabular}{llllrrrr}
\hline
Role & period & $\theta_1$ & $\theta_2$ & $N_0/N_1$ & $\Delta_A$ & $\Delta_B$ & $\Delta_C$ \\
\hline
Parrondo reversal & $00111$ & $3\pi/5$ & $39\pi/40$ & $40/60$ & -15.2411 & -7.8196 & +7.2087 \\
Matched non-paradox & $01101$ & $3\pi/5$ & $39\pi/40$ & $40/60$ & -15.2411 & -7.8196 & -8.9579 \\
\hline
\end{tabular}

\caption{Strict temporal-order-matched pair used in the image simulations,
with \(N=2^8\), \(r=100\), initial position \(\delta=127\), initial coin
state \(\ket{+}\), and effect-size threshold \(\tau=5\). The quantities
\(\Delta_A\), \(\Delta_B\), and \(\Delta_C\) are the exact displacements of
the two fixed-coin walks and the combined walk, respectively, relative to
\(\delta=127\). In both regimes both fixed games lose beyond \(\tau\), while
the combined-game displacements have opposite signs. The angle pair, walk
length, period length, coin-use counts \(N_0/N_1\), initialization, and all
image layers are held fixed, and only the order of the coin uses changes.}
\label{tab:selected-parameters}
\end{table*}

\section{Results and discussion}
Because the keyed walk alternates between coin operators, its reduced walk
dynamics can fall into paradoxical or non-paradoxical regimes. We compare these
regimes under a temporal-order intervention and then use computational-basis
measurements of the coherent transformed state as image diagnostics. We study
a selection of \(64\times64\) pixel images. The matched pair used below is
listed in Table~\ref{tab:selected-parameters}.

\begin{table*}[!htp]
\centering
\begin{tabular}{llrrrrrr}
\hline
Period & role & $\Delta_C$ & $H(\kappa)$ & $p_{(1)}$ & $p_{(2)}$ & $D_{\mathrm{TV}}$ & $F_{\mathrm{state}}$ \\
\hline
$00111$ & selected reversal & +7.2087 & 5.1387 & 0.0903 & 0.0892 & 0.0000 & 1.0000 \\
$01011$ & within threshold & +0.8504 & 4.3364 & 0.1287 & 0.0996 & 0.8709 & 0.0052 \\
$01101$ & matched control & -8.9579 & 4.3445 & 0.1549 & 0.0994 & 0.8733 & 0.0058 \\
$01110$ & cyclic rotation & -19.1478 & 5.0952 & 0.1024 & 0.0956 & 0.2762 & 0.1143 \\
$10011$ & cyclic rotation & +10.2700 & 5.1776 & 0.1281 & 0.0644 & 0.2786 & 0.7685 \\
$10101$ & within threshold & -3.4522 & 4.5133 & 0.1168 & 0.0971 & 0.8536 & 0.0084 \\
$10110$ & within threshold & +0.0444 & 4.6031 & 0.1296 & 0.0803 & 0.8465 & 0.0057 \\
$11001$ & cyclic rotation & -3.5870 & 5.2354 & 0.0899 & 0.0758 & 0.2720 & 0.6830 \\
$11010$ & within threshold & +2.7384 & 4.5120 & 0.1029 & 0.0974 & 0.8463 & 0.0054 \\
$11100$ & cyclic rotation & -14.6561 & 5.1387 & 0.1200 & 0.0737 & 0.2687 & 0.7718 \\
\hline
\end{tabular}

\caption{Exhaustive exact scan of all length-five period words with two
\(\hat C_0\) and three \(\hat C_1\) uses at the selected angles. Every row
has \(N=256\), \(r=100\), \(\delta=127\), the initial coin state \(\ket{+}\),
and hence 40/60 total coin uses. The shared fixed-game displacements are
\(\Delta_A=-15.2411\) and \(\Delta_B=-7.8196\), and outcome roles use
\(\tau=5\). \(H(\kappa)\), in bits, and \(p_{(1)}\) and \(p_{(2)}\) describe the exact
coin-marginalized walk kernel. Total-variation distance
\(D_{\mathrm{TV}}\) and coherent walk-state fidelity
\(F_{\mathrm{state}}\) are measured relative to \(00111\). Rows labeled
cyclic rotation are rotations of that reference word.}
\label{tab:sequence-order-scan}
\end{table*}

Table~\ref{tab:sequence-order-scan} exposes substantial phase and ordering
sensitivity. Even cyclic rotations of \(00111\) span
\(-19.15\leq\Delta_C\leq+10.27\), so the starting phase cannot be discarded
from the control specification. The primary comparison uses \(00111\) and
\(01101\), which are outside one another's cyclic-rotation classes. Holding
all other walk and image settings fixed, temporal order alone changes
\(\Delta_C\) from \(+7.2087\) to \(-8.9579\), crosses both sides of the
predefined threshold, and changes the exact kernel by
\(D_{\mathrm{TV}}=0.8733\). Here order is the intervention, while reversal
status and kernel statistics are both outcomes. The comparison therefore does
not show that reversal itself causes better image scrambling. Because the angle
point and contrast were selected by deterministic screening, it is an
existence result at that point rather than an estimate of an average ordering
effect. The matched classification is also not peculiar to the displayed
\(\tau=5\): all four required strict inequalities retain their roles for any
\(0<\tau<7.20867\).

The displayed image assigns the most frequent observed color to each
coordinate. At 250000 total shots, a coordinate receives only about
\(250000/4096\simeq61\) observations distributed over 256 colors. This modal
estimator can therefore be unstable even though every underlying conditional
probability is generated by a fixed unitary. The figures and conventional
image metrics in the next two subsections describe one seeded, fixed-budget
visualization only. They are not estimators of state-level security or of the
exact per-coordinate mode without a convergence analysis. The exact
conditional distributions and the size of this sampling effect are analyzed
in Section~\ref{sec:per-layer}.

\subsection{Case I: matched non-paradoxical order}
\begin{figure*}[!htp]
\centering
\includegraphics[width=\textwidth]{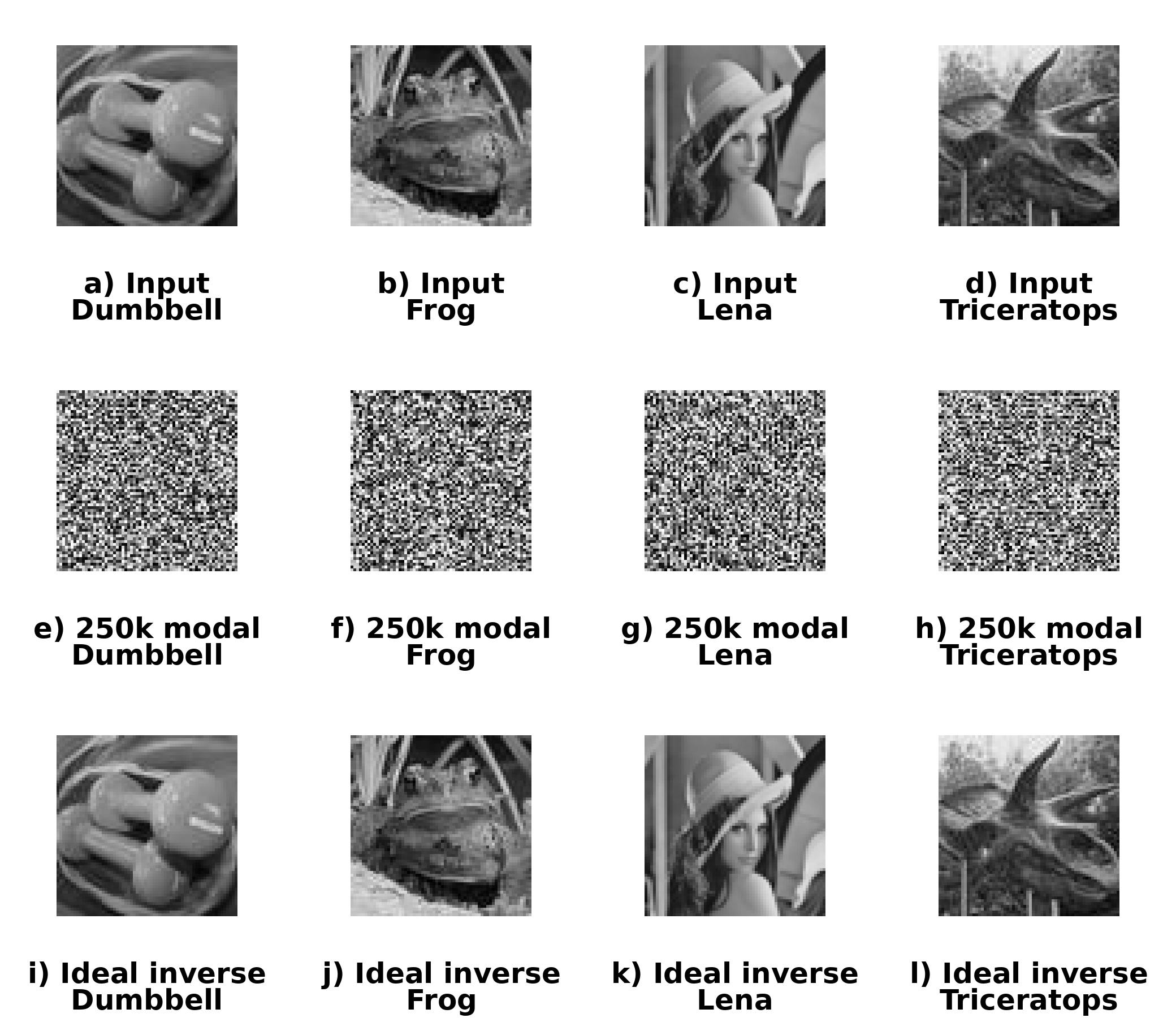}
\caption{Fixed-budget measurement visualizations for the matched
non-paradoxical ordering \(01101\). The first row shows the original images,
the second row
shows 250000-shot modal reconstructions (about 61 observations per coordinate),
and the third row is the analytic ideal-inversion reference
\(U_K^\dagger U_K=\mathbbm{1}\), rendered directly from the input rather than
sampled as a second experiment. The modal reconstructions are not exact
per-coordinate modes (see Section~\ref{sec:per-layer}).}
\label{fig:no-paradox-examples}
\end{figure*}

In Fig.~\ref{fig:no-paradox-examples} we show diagnostic reconstructions for a
matched ordering that does not exhibit the Parrondo paradox. The second row is
obtained by seeded sampling from the exact joint measurement distribution of
independently prepared scrambled states.
The third row shows the analytic ideal-inversion reference
\(U_K^\dagger U_K=\mathbbm{1}\).
The walk parameters are the matched non-paradoxical row of
Table~\ref{tab:selected-parameters}, \(m=01101\,01101\ldots\),
\(\theta_1=3\pi/5\), and \(\theta_2=39\pi/40\), with \(N=2^8\), \(r=100\),
and \(\delta=127\): both fixed games and the combined game lose with margin
\(\tau\).

The modal row appears visually decorrelated at this shot budget. No separate
entropy, correlation, NPCR, or UACI evidence is attached to that appearance,
since all four are nonlinear functions of the same underresolved
per-coordinate argmax. Section~\ref{sec:per-layer} instead evaluates the raw
conditional distributions and exact infinite-shot modes.

\subsection{Case II: Parrondo-reversal order}
We now study a keyed walk whose reduced dynamics satisfies the stated
Parrondo reversal criterion, using the paradoxical row of
Table~\ref{tab:selected-parameters}
(\(m=00111\,00111\ldots\), \(\theta_1=3\pi/5\), \(\theta_2=39\pi/40\)).
The exact displacements are
\(\Delta_A=-15.24\) and \(\Delta_B=-7.82\) for the two fixed-coin games, while
the combined sequence yields \(\Delta_C=+7.21\): both fixed games lose with a
margin above the effect-size threshold, yet the deterministic \(00111\)
switching pattern between the same two coins wins. Its matched \(01101\)
ordering loses with \(\Delta_C=-8.96\).
Figure~\ref{fig:walk-displacement}
illustrates the mechanism with exact statevector quantities. Panel (a) shows the
displacements as a function of the number of steps: the sign structure
\(\Delta_A,\Delta_B<0<\Delta_C\) holds at every integer step count
\(3\le r\le125\) and exceeds the effect-size threshold for
\(72\le r\le125\) (with a single marginal dip,
\(\Delta_C=4.91\), at \(r=71\)). The sign reversal thus persists over a broad
tested step window rather than occurring only at \(r=100\).
The displacements are shown only for \(r\le125\), because the displacement
observable itself is meaningful only there: starting from \(\delta=127\), the
ballistic components reach the boundary of the cycle at \(r\approx N/2=128\),
and once they wrap around, the linear mean \(\mathbb{E}[X_r]-\delta\) no longer
measures drift, a left-moving peak that wraps past \(x=0\) reappears at large
\(x\) and is counted as a large positive displacement. Game outcomes beyond
\(r\approx N/2\) are therefore not defined by this criterion, and walk lengths
used as controls should be kept below \(N/2\), as \(r=100\) is. Panel (b)
shows the corresponding position distributions after \(r=100\) steps:
\(00111\) redistributes the interference pattern so that its mean lies on the
winning side, whereas \(01101\) and both fixed coins lie on the losing side.
Because the
coin rotations and the conditional shift do not commute, the combined evolution
\(W_m=U_{m_r}\cdots U_{m_1}\), with
\(U_b=S(\hat C_b\otimes\mathbbm{1})\), is a temporally ordered product rather
than a convex mixture of the two fixed-coin evolutions. The sequence controls
the relative phases accumulated by
the left- and right-moving components, in analogy with the two-coin Parrondo
games on the line~\cite{rajendran2018implementing,Lai2020}. For the
phaseless rotation coins used here, the plain alternation \(m=0101\ldots\)
produces no threshold-qualified paradox anywhere on the sweep grid. Among the
four tested presets, qualified rows occur only for \(001\) and \(00111\),
and this finite screen does not exclude other sequences or off-grid angles.
Deterministic switching between two losing quantum-walk games was introduced
with phased \(SU(2)\) coins on the line~\cite{chandrashekar2011parrondo}, and
subsequent work found such two-coin advantages to be transient: Flitney mapped
the temporary forward drift of \(AB\)-, \(ABB\)-, and \(ABBB\)-type sequences
of two phase-biased coins, vanishing at long times~\cite{flitney2012}, and
Rajendran and Benjamin reported that for phased coins \(ABB\)- and
\(ABBB\)-type sequences lose asymptotically, with a true paradox requiring a
three-state coin or a second coin
qubit~\cite{rajendran2018playing,rajendran2018implementing}. Asymptotic
paradoxes with \emph{real} coins have been obtained by making the coin
inhomogeneous, either in time, through linearly chirped coin
angles~\cite{pires2020timedependent}, or in space, through site-dependent
coins~\cite{mittal2024inhomogeneous}. The regime reported here differs from
these examples: the two coin operators are static phaseless rotations and
spatially homogeneous, but they are switched periodically in time, and the
walk is launched from the coin state \(\ket{+}\). To test whether the reversal
disappears shortly beyond the \(r\le125\) color-register window, we simulated
the same dynamics on an \(N=2^{13}\) cycle through \(r=4000\), before
wrap-around (\texttt{src/asymptotic\_check.py}). At \(r=4000\), the observed
ratios \(\Delta(r)/r\) are \(+0.0647\) for \(00111\), \(-0.0806\) for its
matched \(01101\) control, and \(-0.1507\) and \(-0.0726\) for the fixed
games. These ratios change only slightly between \(r=1000\) and \(r=4000\).
This finite-size, finite-time calculation demonstrates a long-time pre-wrap
velocity reversal and a persistent sign contrast between the matched
orderings over the tested range, but it does not establish the existence of an
\(r\to\infty\) asymptotic velocity.
On the \(2^q\)-cycle of the color register, the operationally relevant setting
remains the finite window \(r<N/2\), within which the linear displacement
measures drift.

In measurement-based or
semiclassical uses of quantum walks, biased walk distributions may directly
affect the resulting mask or sampled image~\cite{abd2019encryption,Lai2020}.
Here the walk is embedded as one layer of a coherent unitary pipeline, and the
question is how this selected walk kernel appears in fixed-budget and exact
diagnostics of the transformed state.

\begin{figure*}[!htp]
\centering
\includegraphics[width=\textwidth]{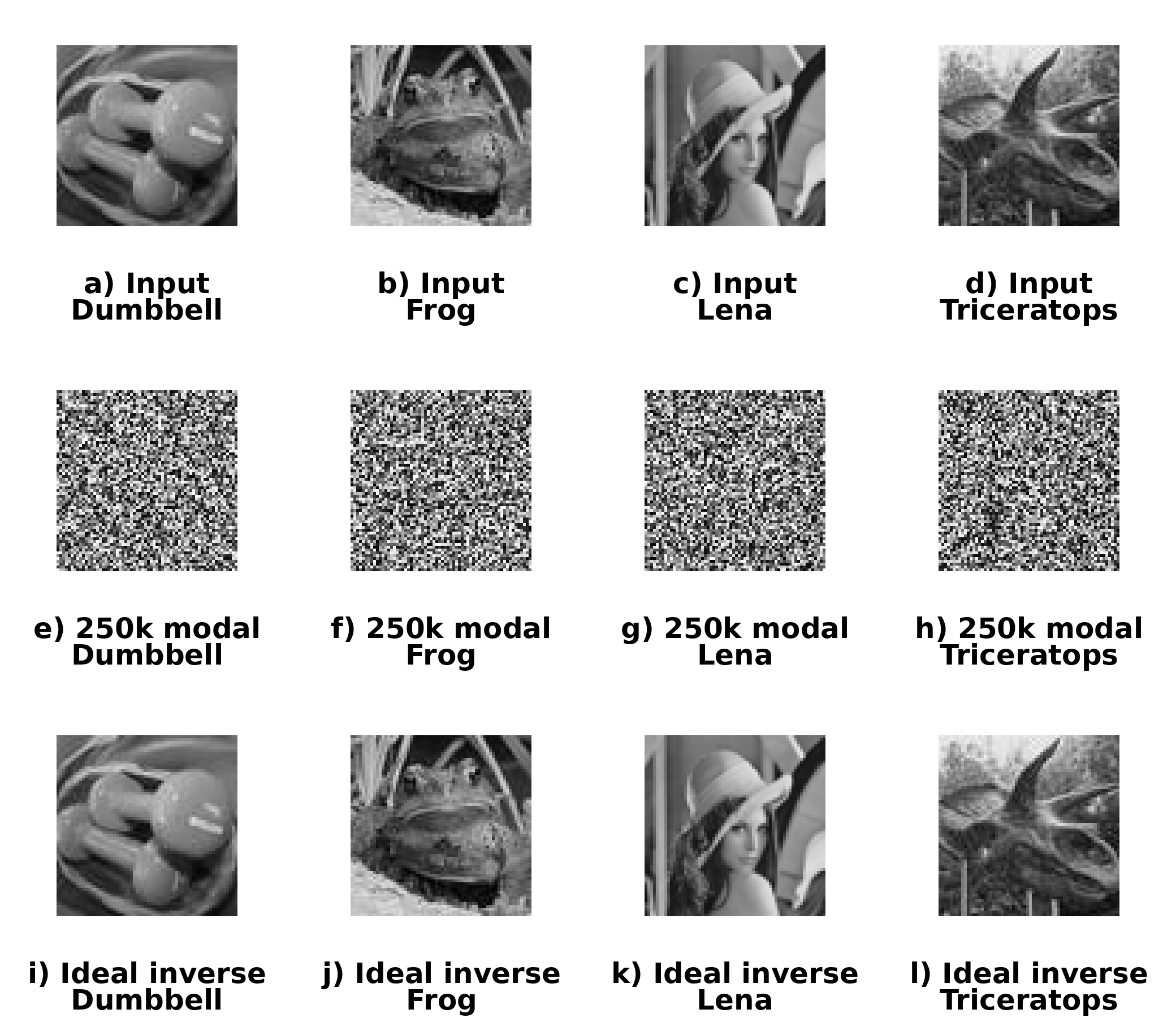}
\caption{Fixed-budget measurement visualizations for the Parrondo-reversal
ordering \(00111\). The first row shows the original images, the second row shows
250000-shot modal reconstructions (about 61 observations per coordinate), and
the third row is the analytic ideal-inversion reference
\(U_K^\dagger U_K=\mathbbm{1}\), rendered directly from the input rather than
sampled as a second experiment. The modal reconstructions are not exact
per-coordinate modes (see Section~\ref{sec:per-layer}).}
\label{fig:paradox-examples}
\end{figure*}

As in the non-paradoxical row, the fixed-budget modal reconstruction has no
obvious image-like pattern. We treat this only as a visualization of the
underresolved estimator. The exact analysis below separates the pooled color
marginal and infinite-shot modal image from finite-shot argmax noise and does
not treat the co-occurring reversal status and kernel differences as evidence
that the reversal causes an image-scrambling advantage.

\subsection{Exact conditional and per-layer diagnostics}
\label{sec:per-layer}

The finite-shot modal images can be replaced by exact quantities because the
walk is translation covariant on the \(2^q\)-cycle. Let \(p=(y,x)\) be an
input address and \(z=\pi(p)\) its address after the fixed public diffusion
permutation. At fixed \(z\), the public position--color circuit XORs a known
byte \(g(z)\) into the color. The walk starts at
\begin{equation}
    a_z=c_p\oplus g(z)\oplus\delta .
\end{equation}
Let \(\kappa(d)\) be the exact coin-marginalized displacement distribution of
the selected coin-and-shift block initialized at displacement zero. Its
dependence on \(m,r,\theta_1,\theta_2\) is implicit, and the XOR offset is
omitted because it is already included in \(a_z\). Then
\begin{equation}
    P_K(J=j\mid Z=z)
    =\kappa\!\left((j-a_z)\bmod 2^q\right).
    \label{eq:conditional-shift}
\end{equation}
Every coordinate thus has the same conditional shape, translated by its
mixed color. In particular,
\begin{equation}
    H(J\mid Z)=H(\kappa),\qquad
    I(Z;J)=H(J)-H(\kappa).
\end{equation}
These quantities refer to the fixed Lena state with uniformly measured
coordinates. They are state diagnostics rather than a general
plaintext-security measure.

\begin{table*}[!htp]
\begin{tabular}{lrrrrrrrr}
\hline
Regime & $H(\kappa)$ & $p_{(1)}$ & $p_{(2)}$ & $E[n_{(1)}]$ & $H(J)_W$ & $I(Z;J)_W$ & $H(J)_{\mathrm{full}}$ & $I(Z;J)_{\mathrm{full}}$ \\
\hline
paradox & 5.1387 & 0.0903 & 0.0892 & 5.51 & 7.9538 & 2.8150 & 7.9982 & 2.8594 \\
non-paradox & 4.3445 & 0.1549 & 0.0994 & 9.46 & 7.7539 & 3.4093 & 7.9974 & 3.6528 \\
\hline
\end{tabular}

\caption{Exact conditional and pooled-color diagnostics for Lena.
\(\kappa\) is the common conditional walk kernel, \(p_{(1)}\) and
\(p_{(2)}\) are its two largest probabilities, and \(E[n_{(1)}]\) is the
expected count of the most likely color at the manuscript budget of 250000
shots. \(H(J)_W\) and \(H(J)_{\mathrm{full}}\) are exact pooled color
entropies for W alone and D+M+W. The corresponding \(I(Z;J)\) values use the
jointly measured coordinate and color. All entropies and mutual informations
are in bits. High pooled entropy does not imply a
uniform conditional distribution. Here and in subsequent generated tables,
non-paradox denotes the matched \(01101\) ordering.}
\label{tab:exact-walk}
\end{table*}

Table~\ref{tab:exact-walk} explains the modal instability. In the paradoxical
row, the leading probabilities are \(0.09030\) and \(0.08921\), giving only
5.51 expected observations of the leading value at each coordinate. In the
matched non-paradoxical row the corresponding expectation is 9.46. The exact modal
color is
\begin{equation}
    j^*(z)=(a_z+d^*)\bmod 2^q,\qquad
    d^*=\operatorname*{arg\,max}_d\kappa(d).
\end{equation}
For W alone this is a bijective byte relabeling
\(c\mapsto((c\oplus\delta)+d^*)\bmod256\), so it preserves the original
histogram entropy exactly. Both selected rows give
\(H_{W,\mathrm{mode}}=7.5122\). Finite-shot modal estimates need not share
that value or even recover the exact modal color.

\begin{table*}[!htp]
\centering
\begin{tabular}{lrrrrr}
\hline
Regime & total shots & shots/coordinate & mode agreement (\%) & $H_{W,\mathrm{mode}}$ & $H_{\mathrm{full,mode}}$ \\
\hline
paradox & 250000 & 61.0 & 30.37 $\pm$ 0.76 & 7.9001 $\pm$ 0.0060 & 7.9530 $\pm$ 0.0039 \\
paradox & 1000000 & 244.1 & 42.28 $\pm$ 0.85 & 7.8945 $\pm$ 0.0059 & 7.9540 $\pm$ 0.0031 \\
paradox & 10000000 & 2441.4 & 54.99 $\pm$ 1.08 & 7.8880 $\pm$ 0.0064 & 7.9530 $\pm$ 0.0024 \\
paradox & 100000000 & 24414.1 & 65.37 $\pm$ 0.84 & 7.8662 $\pm$ 0.0062 & 7.9541 $\pm$ 0.0026 \\
paradox & 1000000000 & 244140.6 & 89.64 $\pm$ 0.59 & 7.6895 $\pm$ 0.0132 & 7.9575 $\pm$ 0.0021 \\
non-paradox & 250000 & 61.0 & 69.15 $\pm$ 0.89 & 7.5929 $\pm$ 0.0077 & 7.9563 $\pm$ 0.0028 \\
non-paradox & 1000000 & 244.1 & 93.93 $\pm$ 0.40 & 7.5286 $\pm$ 0.0039 & 7.9590 $\pm$ 0.0018 \\
non-paradox & 10000000 & 2441.4 & 100.00 $\pm$ 0.00 & 7.5122 $\pm$ 0.0000 & 7.9596 $\pm$ 0.0000 \\
non-paradox & 100000000 & 24414.1 & 100.00 $\pm$ 0.00 & 7.5122 $\pm$ 0.0000 & 7.9596 $\pm$ 0.0000 \\
non-paradox & 1000000000 & 244140.6 & 100.00 $\pm$ 0.00 & 7.5122 $\pm$ 0.0000 & 7.9596 $\pm$ 0.0000 \\
\hline
\end{tabular}

\caption{Convergence of the finite-shot modal estimator for Lena under direct
multinomial sampling from the exact conditional kernels. Entries are means
\(\pm\) sample standard deviations over 16 deterministically seeded
replicates. Mode agreement is the fraction of coordinates assigned the exact
kernel mode, and entropies are in bits. Non-paradox denotes the matched \(01101\)
ordering.}
\label{tab:shot-convergence}
\end{table*}

Table~\ref{tab:shot-convergence} makes the convergence problem explicit.
At 250000 total shots, 16 replicates give modal agreement (mean \(\pm\) sample
standard deviation over replicates)
\(30.37\%\pm0.76\%\) and \(69.15\%\pm0.89\%\), respectively. At ten million
shots the values are \(54.99\%\pm1.08\%\) and \(100.00\%\pm0.00\%\). Because
the paradoxical kernel has nearly tied peaks, even \(10^9\) shots gives only
\(89.64\%\pm0.59\%\). The finite-shot apparent W-only entropy increase and
the poorly resolved modal image in the reversal row are consequently
artifacts of the undersampled argmax estimator. The matched control, whose
kernel has a wider peak gap, converges much faster.
The exact correlations also differ under the temporal-order intervention.
That identifies order as the source of the kernel-dependent difference at
this point, but it does not identify reversal as its cause: reversal status
and the correlations are simultaneous outcomes of the reordered product.

\begin{table*}[!htp]
\begin{tabular}{llrrrr}
\hline
Exact diagnostic & Regime & $H$ & $C_H$ & $C_V$ & $C_D$ \\
\hline
D+M & --- & 7.9596 & -0.1935 & 0.0857 & -0.0718 \\
W exact mode & paradox & 7.5122 & 0.6361 & 0.7431 & 0.6148 \\
D+M+W exact mode & paradox & 7.9596 & -0.0382 & -0.0233 & 0.0349 \\
W exact mode & non-paradox & 7.5122 & 0.6055 & 0.7525 & 0.5348 \\
D+M+W exact mode & non-paradox & 7.9596 & -0.1065 & 0.0092 & -0.0048 \\
\hline
\end{tabular}

\caption{Exact infinite-shot modal diagnostics for Lena. D+M is the
deterministic image after the two fixed public layers. W and D+M+W use the
exact per-coordinate mode of Eq.~\eqref{eq:conditional-shift}, with no
measurement sampling. \(H\) is in bits. Correlations follow the fixed
2000-pair subsampling protocol, and the image and its modal values are
otherwise exact.}
\label{tab:exact-layers}
\end{table*}

The exact full-pipeline modal images retain the high entropy introduced by the
public D+M map and have small adjacent-pixel correlations for both selected
walks (Table~\ref{tab:exact-layers}). The exact pooled color entropies are even
closer to eight, \(7.9982\) and \(7.9974\). This establishes a property of the
implemented public composition, but not a Parrondo advantage. Although the
rows are strictly matched except for temporal order, the walk changes both
the common kernel and a global modal byte relabeling, and pooled entropy is
nearly saturated in both cases. The information values \(2.8594\) and
\(3.6528\) bits for the reversal and matched non-paradoxical orderings,
respectively, further show why a nearly uniform pooled histogram must not be
confused with uniformity conditional on a measured coordinate. Their
difference is descriptive for this post-selected pair, not evidence of a
general reversal advantage.

\subsubsection{Repeated-copy computational-basis access}
\label{sec:repeated-copy}

The following observation assumes multi-copy access through repeated
transmissions, repeated use of the state-preparation procedure, or an oracle.
It does not follow from possession of one unknown state, which cannot be
cloned. Under multi-copy access an observer can group measured colors by
coordinate \(z\). Equation~\eqref{eq:conditional-shift} shows that all of the
resulting histograms are cyclic shifts of one unknown template. Cyclic
cross-correlation recovers their relative shifts whenever
\(\kappa\) has no translational symmetry and the sample budget is adequate.
The two selected kernels satisfy this condition: their autocorrelations have
a unique maximum at zero shift, with the next peak only \(0.639\) and \(0.723\)
of that maximum.

This alignment does not require prior knowledge of \(\kappa\). Once the
relative \(a_z\) are known, the unknown global cyclic alignment and the
unknown \(q\)-bit XOR offset leave at most \(2^{2q}=2^{16}\) candidate images
for \(q=8\). If the experiment's fixed \(\delta=127\) is treated as public,
only the \(2^q\) global alignments remain, before using a known pixel,
candidate-template test, or natural-image prior. We do not quantify the
attack's sample complexity here.
The alignment nevertheless shows that the deterministic map is not a
confidentiality construction under repeated-copy computational-basis access. A
security-oriented variant would require per-message randomization and a keyed
non-covariant layer that breaks the common-template alignment.

\subsection{Key sensitivity of the coherent map}\label{sec:key-sensitivity}

All variable control dependence in the implemented map lies in the walk
layer. A natural diagnostic of coherent inversion is the recovery fidelity
under perturbed control parameters,
\begin{equation}
    F(K,K') = \left|\bra{\Psi_0} U_{K'}^\dagger U_K \ket{\Psi_0}\right|^2,
\end{equation}
i.e., the fidelity between the original state and the state recovered with a
perturbed key \(K'\). Because the diffusion and mixing layers are
key-independent and map computational-basis states to computational-basis
states, \(F\) reduces exactly to an average of walk-layer amplitudes over the
distribution \(p(\tilde c)\) of mixed color values of the image,
\begin{equation}
    \bra{\Psi_0} U_{K'}^\dagger U_K \ket{\Psi_0}
    = \sum_{\tilde c} p(\tilde c)\,
    \bra{0,\tilde c} W_{K'}^\dagger W_K \ket{0,\tilde c},
    \label{eq:fidelity-reduction}
\end{equation}
which we evaluate from exact statevectors of the implemented walk circuits.
We verified Eq.~\eqref{eq:fidelity-reduction} against a full 21-qubit
statevector simulation of \(U_{K'}^\dagger U_K\) on the Lena image, with
agreement at the \(10^{-15}\) level.

\begin{figure}[!htp]
\centering
\includegraphics[width=\columnwidth]{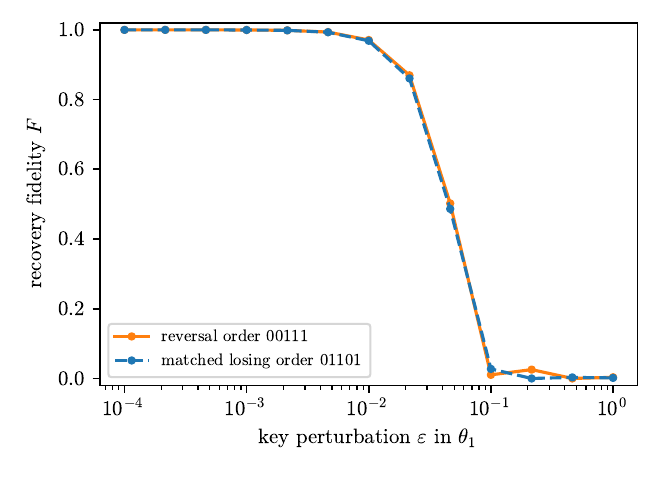}
\caption{Exact recovery fidelity \(F\) of the Lena image state under
inversion with a perturbed coin angle \(\theta_1+\varepsilon\), locally around
the two matched temporal orderings.}
\label{fig:key-sensitivity}
\end{figure}

Figure~\ref{fig:key-sensitivity} shows \(F\) as a function of a perturbation
\(\varepsilon\) of the coin angle \(\theta_1\). The fidelity is flat near one
for very small errors (\(F>0.999\) at \(\varepsilon=10^{-3}\)), falls through
\(1/2\) near \(\varepsilon\approx4.6\times10^{-2}\), and is near zero by
\(\varepsilon=10^{-1}\) in both orderings. The perturbed coin acts in 40 of
the \(r=100\) walk steps in each case, so coin errors accumulate coherently,
and the small difference between the curves comes only from their temporal
order.
These values measure local recovery sensitivity around the two reference rows
only. In particular, the \(F=1/2\) crossover is not an estimate of angular key
resolution, key entropy, or parameter-recovery complexity. Discrete control
perturbations are summarized in Table~\ref{tab:discrete-fidelity}. Both
admit exact closed forms, which the numerics reproduce to all printed digits
and which show that these perturbations probe the structure of the circuit
rather than the walk dynamics. Because the offset enters as the bitwise map
\(\ket{c}\mapsto\ket{c\oplus\delta}\) at the walk input and every other keyed
element is unchanged,
\(W_{K'}^\dagger W_K=\1_{\mathrm{coin}}\otimes X^{(\delta\oplus\delta')}\),
where \(X^{(\nu)}\) flips the color bits set in
\(\nu=\delta\oplus\delta'\). Every diagonal matrix element
\(\bra{\tilde c}X^{(\nu)}\ket{\tilde c}\) with \(\nu\neq0\)
vanishes, so \(F=0\) identically for \emph{any} wrong offset, in both regimes
and for all coin parameters: the orthogonality reported in the table is a
structural identity of the construction, not an avalanche effect. A single
flipped bit of the coin sequence likewise has a closed form. The flip
replaces \(\hat C_0\) by \(\hat C_1\) (or vice versa) at one step, so
\(W_{K'}^\dagger W_K\) reduces to the single inserted coin rotation
\(\hat C_1^\dagger\hat C_0\), a rotation by \(\theta_1-\theta_2\), conjugated
by the walk evolution preceding the flipped step. That evolution is a real
matrix in the computational basis (the coins, the conditional shift, the
offset, and the coin Hadamard are all real), so it maps
\(\ket{0,\tilde c}\) to a real vector, and for any real unit vector
\(\ket{\psi}\) and real coin rotation \(R(\varphi)\) one has
\(\bra{\psi}(R(\varphi)\otimes\1)\ket{\psi}=\cos\varphi\). Hence
\begin{equation}
    F_{\mathrm{flip}} = \cos^2(\theta_1-\theta_2),
    \label{eq:flip-fidelity}
\end{equation}
independently of the flip position, the number of steps, the offset, and the
image. Both matched rows use the same angles, so both table values are exactly
\(\cos^2(3\pi/8)=0.146447\). Any ordering with the same angle separation
responds identically to a single-bit sequence error, so this diagnostic
contains no dependence on Parrondo-reversal status.

Here NPCR is the fraction of pixels whose reconstructed and original values
differ, and UACI is the mean absolute gray-level difference normalized by
255. These definitions apply only to the sampled image estimator in this
table, not to the coherent state itself.

The sampled wrong-key diagnostics in Table~\ref{tab:wrong-key-npcr} show the
image-level consequence together with its limits. With the correct key
the reconstruction is pixel-perfect, and a coin-angle error of \(10^{-1}\)
changes essentially every pixel. The offset row, by contrast, must not be
read as scrambling. Since
\(W_{K'}^\dagger W_K=\1_{\mathrm{coin}}\otimes X^{(\delta\oplus\delta')}\)
and the color-bit flips commute with the diffusion layer, which acts only on
the coordinates, and with the mixing layer, whose CNOT and Toffoli gates
target color bits and are controlled by coordinates, the wrong-offset
reconstruction is the deterministic pixelwise map
\(I\mapsto I\oplus(\delta\oplus\delta')\): an invertible relabeling of the
gray values that leaks the plaintext completely. For the tested
\(\delta'=\delta+1\) the mask is \(255\) and the reconstruction is exactly
the photographic negative \(255-I\), whose adjacent-pixel correlations equal
those of the original image. The table entries match this description: NPCR is
\(100\%\) only because every pixel value changes deterministically, and the
UACI of \(34.0\%\) is exactly the UACI between the Lena image and its
negative, not the \(29.4\%\) baseline of a reconstruction unrelated to the
original. The offset component of the key thus provides no concealment
at the measurement level, even though the corresponding coherent states are
exactly orthogonal. For small perturbations the mode-based reconstruction is
instead far more forgiving than the coherent fidelity: at
\(\varepsilon=10^{-2}\)
(\(F\approx0.97\)) and even for the flipped sequence bit in the paradoxical
regime (\(F\approx0.15\)), the per-pixel sampled plurality over \(\sim61\)
samples still recovers most pixels, because the residual amplitude on the
correct color value dominates the mode. Both failure modes point the same way.
Coherent-state fidelity and measured image distance are different quantities:
\(F=0\) coexists with a complete information leak at readout, and
\(F\approx0.15\) with near-perfect recovery. The fidelity of the coherent
state, rather than a measured image distance, is the appropriate figure of
merit for the keyed map, and measured reconstructions are not the
cryptographic object.

\begin{table}[!htp]
\begin{tabular}{lrr}
\hline
Perturbation & $F$ ($00111$) & $F$ ($01101$) \\
\hline
flip one bit of $m$ & 0.146447 & 0.146447 \\
$\delta \to \delta + 1$ & 0.000000 & 0.000000 \\
\hline
\end{tabular}

\caption{Exact recovery fidelity for discrete key perturbations of the Lena
image state, computed via Eq.~\eqref{eq:fidelity-reduction}. Both values
match the closed forms derived in the text: \(F=0\) identically for any
offset change, and \(F_{\mathrm{flip}}=\cos^2(\theta_1-\theta_2)\) of
Eq.~\eqref{eq:flip-fidelity} for a single flipped sequence bit.}
\label{tab:discrete-fidelity}
\end{table}

\begin{table}[!htp]
\begin{tabular}{lrr}
\hline
Perturbation & NPCR (\%) & UACI (\%) \\
\hline
$\varepsilon = 0$ (correct key) & 0.0000 & 0.0000 \\
$\theta_1 + 10^{-3}$ & 0.0000 & 0.0000 \\
$\theta_1 + 10^{-2}$ & 0.0000 & 0.0000 \\
$\theta_1 + 10^{-1}$ & 99.9756 & 7.3024 \\
flip one bit of $m$ & 5.1270 & 1.5765 \\
$\delta + 1$ & 100.0000 & 34.0056 \\
\hline
\end{tabular}

\caption{Sampled wrong-key inversion diagnostics for the Lena image in the
paradoxical regime (250000 shots): NPCR and UACI between the original image
and the mode-based reconstruction after applying \(U_{K'}^\dagger U_K\).}
\label{tab:wrong-key-npcr}
\end{table}

\subsection{Noise robustness of the scrambling/inversion cycle}\label{sec:noise}

We also quantify the sensitivity of the coherent cycle
\(U_K^\dagger U_K\) to gate noise in the keyed walk layer. We attach a
depolarizing channel to every gate of the transpiled walk circuit and its
inverse, with Qiskit's depolarizing parameter \(p_2\) after each two-qubit
gate and \(p_2/10\) after each one-qubit gate, while treating the NEQR
preparation and the Clifford
diffusion/mixing layers as ideal, so that the curves isolate the walk layer,
which dominates the two-qubit gate count of the scrambling stage
(Table~\ref{tab:resources}). For a \(k\)-qubit gate, Aer uses
\(\mathcal{D}^{(k)}_p(\rho)=(1-p)\rho+pI_{2^k}/2^k\), so \(p\) is the
channel parameter rather than the total probability of a nonidentity Pauli.
Writing \(\mathcal{N}\) for the noisy
scrambling/inversion cycle of the walk layer, we compute from 9-qubit
density-matrix simulations the average basis-state recovery fidelity over
the mixed-color distribution of the Lena image,
\begin{equation}
    \bar F = \sum_{\tilde c} p(\tilde c)\,
    \bra{0,\tilde c}\mathcal{N}\!\left(\ketbra{0,\tilde c}{0,\tilde c}\right)
    \ket{0,\tilde c},
    \label{eq:noise-fidelity}
\end{equation}
together with the pixel error rate of the recovered image. This ensemble
average is not the fidelity of the coherent image state. For the particular
gate-local depolarizing model used here, however, it is an upper bound. The
channel has a random-unitary decomposition
\(\mathcal{N}(\rho)=\sum_s q_sV_s\rho V_s^\dagger\), where each \(V_s\)
is the circuit with one definite pattern of Pauli errors. Writing
\(v_{s,\tilde c}=\bra{0,\tilde c}V_s\ket{0,\tilde c}\), branchwise convexity
gives
\begin{equation}
    F_{\mathrm{coh}}=
    \sum_s q_s\left|\sum_{\tilde c}p(\tilde c)v_{s,\tilde c}\right|^2
    \leq
    \sum_s q_s\sum_{\tilde c}p(\tilde c)|v_{s,\tilde c}|^2
    =\bar F .
\end{equation}
This inequality is specific to the random-unitary noise model and need not
hold for an arbitrary nonunitary channel. The pixel error rate, by contrast,
is exact as computed:
measuring the coordinate registers selects a single pixel, so the color
statistics of the recovered image depend only on the diagonal blocks
appearing in Eq.~\eqref{eq:noise-fidelity}.

\begin{figure*}[!htp]
\centering
\includegraphics[width=\textwidth]{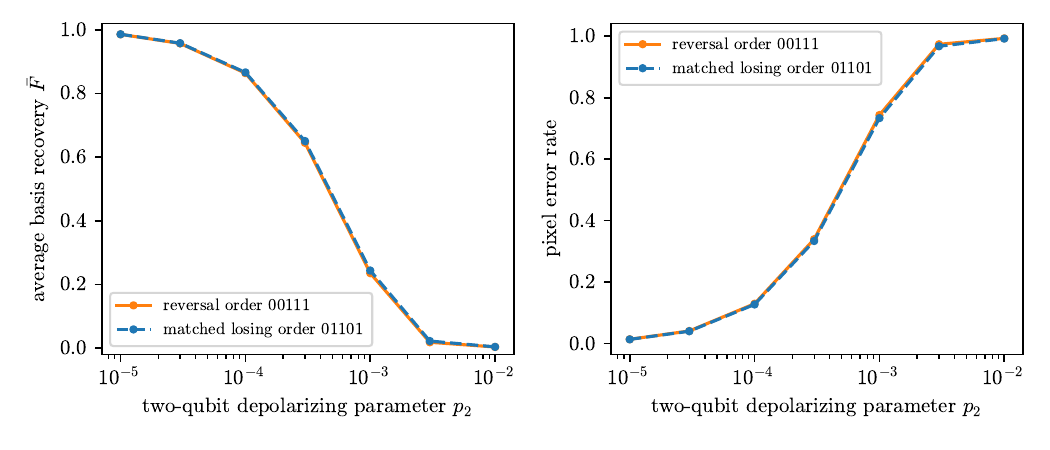}
\caption{Effect of depolarizing gate noise in the walk layer on the
scrambling/inversion cycle for the Lena image state: (left) average
basis-state recovery fidelity \(\bar F\) of Eq.~\eqref{eq:noise-fidelity},
an upper bound on the recovery fidelity of the full image state for this
random-unitary depolarizing model, and (right)
pixel error rate of the recovered image (exact), as functions of the
two-qubit depolarizing parameter \(p_2\) (one-qubit parameter \(p_2/10\)), for the
reversal and matched losing orderings. Both quantities are computed from
exact density-matrix simulations of the noisy walk cycle.}
\label{fig:noise-robustness}
\end{figure*}

The results are shown in Fig.~\ref{fig:noise-robustness}. The walk cycle
contains roughly \(1.5\times10^3\) noisy two-qubit gate applications, so the
average basis recovery \(\bar F\) decays on the scale \(p_2\sim10^{-3}\).
Across the two orderings, \(\bar F\) ranges from \(0.9854\) to \(0.9857\) at
\(p_2=10^{-5}\), from \(0.8636\) to \(0.8659\) at \(10^{-4}\), and from
\(0.2353\) to \(0.2435\) at \(10^{-3}\). The corresponding pixel-error ranges
are \(1.35\%\)--\(1.38\%\), \(12.72\%\)--\(12.96\%\), and
\(73.37\%\)--\(74.36\%\). The two selected orderings behave very similarly,
since their circuits have the same angles, gate counts, and topology and
differ only in the order of the same coin operations. Under this idealized model, keeping
the average basis recovery close to one requires depolarizing parameters in the
\(10^{-5}\)--\(10^{-4}\) range. The curves do not determine the full coherent
fidelity or a fault-tolerance threshold, and they exclude noise in the
dominant NEQR preparation. Computing the full-state channel fidelity and a
hardware-specific resource analysis remains future work.

\subsection{Limits of comparison with classical cipher-image metrics}
\label{sec:comparison}

Published sampled-mask designs such as
Refs.~\cite{liu2022dtqwaes,zhao2022rubik,zhang2022zigzag,alblehai2025cascading,liang2023lstm}
report metrics of a transmitted classical cipher image. Their NPCR and UACI
are normally differential quantities between two cipher images whose
plaintexts differ in one pixel. Here the transformed object is a coherent
state, while the displayed image is a nonlinear finite-shot estimator, and
our NPCR/UACI compare that estimator with its plaintext. These quantities
share neither a statistical object nor a null model.

We do not rank the present values against the published
cipher-image numbers or claim statistical equivalence. Entropy, correlation,
NPCR, and UACI are retained only as internal descriptions of the fixed-budget
visualizations and the explicitly labeled exact modal maps. A meaningful
cross-scheme comparison would require the same transmitted object, access
model, plaintext ensemble, estimator, image size, and differential
experiment.

\section{Conclusions}

We analyzed a coherent reversible transformation of NEQR image states based on
a discrete-time quantum walk. The classical control vector selects the walk
sequence, angles, length, and color offset, while the implemented diffusion
and position--color circuits are fixed and public. The construction does not
prepare or discard a quantum key image, and inversion acts on the complete
coherent state, including the coin register, before measurement.

On the walk side, exact statevector expectations identify five grid points at
which two negative-drift fixed games combine into a positive-drift periodically
switched game beyond the deterministic effect-size threshold. At the strongest
row of the four-preset sweep, an exhaustive same-composition scan gives a strict temporal-order
control: with identical angles, period length, initialization, and 40/60 coin
uses, \(00111\) gives \(\Delta_C=+7.2087\), whereas \(01101\) gives
\(\Delta_C=-8.9579\). Their velocity ratios remain \(+0.0647\) and
\(-0.0806\), respectively, through \(r=4000\) on a larger cycle before
wrap-around. This is a long-time finite-window result, not a proof of an
\(r\to\infty\) asymptotic velocity. The full ten-word scan also shows that
cyclic starting phase matters and must be included in the ordered control.
Plain sign tests on sampled displacements are unreliable: at realistic shot budgets
they flag ``paradoxes'' at chance level even for parameter sets whose games
are physically identical, so a margin criterion on exact expectations (or a
properly powered statistical test) is necessary. Among the four tested
sequence presets, threshold-qualified rows use periods \(001\) or \(00111\),
while plain alternation produces none on the grid. The finite screen does not
exclude other sequences or off-grid angles.

For the image application, the exact translation-covariance analysis changes
the interpretation of the sampled figures. Every conditional color
distribution is a shift of one kernel, and the leading paradoxical kernel
probabilities are nearly tied. At 250000 shots (about 61 observations per
coordinate), 16 replicates recover its exact W-only mode at only
\(30.37\%\pm0.76\%\) of Lena coordinates, compared with
\(69.15\%\pm0.89\%\) for the matched control. The exact W-only mode is a byte
permutation and preserves the original entropy \(7.5122\), and the finite-shot
apparent increase is sampling noise. The full D+M+W composition, whose D+M
layers are fixed and public,
has exact pooled color entropies \(7.9982\) and \(7.9974\), and small exact
modal correlations in both selected rows. Temporal order is the sole
intervention in this comparison, but reversal status and all kernel-dependent
image statistics are joint outcomes. The matched result therefore isolates an
order effect without establishing a causal image-scrambling advantage from
Parrondo reversal.

The same common-kernel structure permits relative-shift recovery by
conditional-histogram alignment under repeated-copy computational-basis
access. The implemented deterministic map is therefore a reversible coherent
transformation benchmark, not a confidentiality protocol. The exact
wrong-control analysis remains useful locally: an offset change gives
orthogonal coherent states while leaking a bitwise relabeling at readout, and
a single sequence-bit flip gives
\(F=\cos^2(\theta_1-\theta_2)\). The angular fidelity crossover is a local
sensitivity measure, not key entropy. Under depolarizing noise, the reported
\(\bar F\) is an average basis-recovery upper bound on full-state fidelity,
and the full coherence cost has not been computed.

The resource estimates show that the efficient QFT-based DTQW layer is
not the dominant cost for arbitrary classical inputs. Exact NEQR
loading contributes the largest high-level gate count instead. Practical use
depends on how the image state is supplied and on the cost of preparation.
The principal open tasks are an analytic Floquet characterization
of the long-time velocity, phase-averaged or preregistered broader sequence
ensembles, full coherent channel fidelity under noise, and randomized
non-covariant constructions evaluated under an explicit security definition.

\section*{Acknowledgments}
The author would like to thank Markus Grassl for insightful comments.

\section*{Statements and Declarations}

\subsection*{Funding}
This work was supported by the Institute of Theoretical and Applied
Informatics, Polish Academy of Sciences, through internal project
IITIS/BW/04/25.

\subsection*{Competing interests}
The author declares no competing interests.

\subsection*{Data availability}
The numerical data underlying the figures and tables are supplied with the
manuscript sources and supplementary files.

\subsection*{Code availability}
The Python/Qiskit sources, pinned environment specification, and launch script
used to regenerate the reported results are supplied with the manuscript
sources and supplementary files.

\subsection*{Author contributions}
The sole author conceived the study, developed the methodology and software,
performed the simulations and analysis, prepared the visualizations, and
wrote and revised the manuscript.
\bibliographystyle{apsrev4-2}
\bibliography{bibliography}

\end{document}